\pdfoutput=1

\documentclass[12pt,epsf]{article}
\usepackage{graphicx}
\newcommand{\be}{\begin{equation}}
\newcommand{\ee}{\end{equation}}
\newcommand{\bea}{\begin{eqnarray}}
\newcommand{\eea}{\end{eqnarray}}
\newcommand{\beas}{\begin{eqnarray*}}
\newcommand{\eeas}{\end{eqnarray*}}
\newcommand{\ba}{\begin{array}}
\newcommand{\ea}{\end{array}}

\newcommand{\nbox}{{\,\lower0.9pt\vbox{\hrule \hbox{\vrule height 0.2 cm \hskip 0.19 cm \vrule height 0.2 cm}\hrule}\,}}

\def\href#1#2{#2}

\begin{document}
\begin{titlepage}
\hfill

\vspace*{20mm}
\begin{center}
{\Large \bf Towards A Holographic Model \\ of Color Superconductivity}

\vspace*{15mm}
\vspace*{1mm}
Pallab Basu${}^{a,b}$, Fernando Nogueira${}^{a}$, Moshe Rozali${}^{a}$,\\ Jared B. Stang${}^{a}$, Mark Van Raamsdonk${}^{a}$ \footnote{e-mails: pallabbasu@gmail.com, nogueira@phas.ubc.ca, rozali@phas.ubc.ca, jstang@phas.ubc.ca, mav@phas.ubc.ca}
\vspace*{1cm}

{\it ${}^{a}$Department of Physics and Astronomy,
University of British Columbia\\
6224 Agricultural Road,
Vancouver, B.C., V6T 1W9, Canada\\
${}$ \\
${}^{b}$
Department of Physics and Astronomy, \\
University of Kentucky, Lexington, KY 40506, USA
}

\vspace*{1cm}
\end{center}

\begin{abstract}
In this note, we discuss the basic elements that should appear in a gravitational system dual to a confining gauge theory displaying color superconductivity at large baryon density. We consider a simple system with these minimal elements, and show that for a range of parameters, the phase structure of this model as a function of temperature and baryon chemical potential exhibits phases that can be identified with confined, deconfined, and color superconducting phases in the dual field theory. We find that the critical temperature at which the superconducting phase disappears is remarkably small  (relative to the chemical potential). This small number arises from the dynamics, and is unrelated to any small parameter in the model that we study.  We discuss similar models which exhibit flavor superconductivity.

\end{abstract}
\vskip 2cm

\begin{center}
\vskip 1cm

\end{center}
\end{titlepage}

\vskip 1cm
\vskip 0.1 in
\noindent

\section{Introduction}

\subsubsection*{Background}

Quantum Chromodynamics is believed to display a rich phase structure at finite temperature and chemical potential, with phase transitions associated with deconfinement, nuclear matter condensation, the breaking of (approximate) flavor symmetries (which are exact in generalizations with equal quark masses and/or massless quarks), and the onset at high density of quark matter phases displaying color superconductivity (for reviews see for example \cite{Alford:2007xm, Rajagopal:2000wf, Stephanov:2007fk,McLerran:2007qj}). However, apart from the regimes of asymptotically large temperature or chemical potential, a direct analytic study of the thermodynamic properties of the theory is not possible.

Even using numerical simulations, only the physics at zero chemical potential is currently accessible, since at finite $\mu$ the Euclidean action becomes complex, and the resulting oscillatory path integral cannot reliably be simulated using standard Monte-Carlo techniques. Current proposals for the phase diagram of QCD and related theories are largely based on qualitative arguments and phenomenological models. While these provide a plausible picture, it is possible that they miss important features of the physics. It would certainly be satisfying to have examples of theories similar to QCD in which the full phase diagram could be explored directly.

\subsubsection*{The holographic approach}

A modern route to understanding properties of strongly coupled gauge theories, that would be otherwise inaccessible, is via the AdS/CFT correspondence, or gauge theory / gravity duality. This suggests that certain quantum field theories (usually called ``holographic theories''), generally with large-rank gauge groups, are equivalent to gravitational systems. By this correspondence, calculations of physical observables in the field theory are mapped to gravitational calculations; in many cases difficult strongly-coupled quantum mechanical calculations in the field theory (such as those required to understand the thermodynamic properties of QCD) are mapped to relatively simple classical gravity calculations. Optimistically, it may then be possible to find a theory qualitatively similar to QCD for which the physics at arbitrary temperature and chemical potential can be understood exactly via simple calculations in a dual gravitational system.

By now, there are well-known examples in gauge-theory / gravity duality for which the field theory shares many of the qualitative features of QCD (see, for example \cite{Sakai:2004cn}). Further, many of these theories have been studied at finite temperature and chemical potential, revealing phase transitions associated with deconfinement, chiral symmetry breaking, meson melting, and the condensation of nuclear matter. However, to date, most of the theories that can be studied reliably using dual gravity calculations have the restriction that the number of flavors is kept fixed in the large $N_c$ limit. In such theories, the physics at large chemical potentials is known to be qualitatively different than in real QCD. For example, at asymptotically large chemical potential, theories with large $N_c$ and fixed $N_f$ are believed to exhibit an inhomogeneous ``chiral density wave'' behavior \cite{Deryagin:1992rw, Shuster:1999tn}, rather than the homogenous quark matter phases predicted for finite $N_c$ and $N_f$. In order to find examples of holographic theories which most closely resemble real QCD at finite chemical potential, one should therefore attempt to find examples of calculable gravitational systems corresponding to theories with finite $N_f/N_c$. This situation presents some technical challenges, as we now review.

In the well-known examples of holographic gauge theories, the addition of flavor fields in the field theory corresponds to adding D-branes on the gravity side \cite{Karch:2002sh}. Quarks correspond to strings which have one endpoint on these D-branes, while mesons correspond to the quantized modes of open strings which begin and end on the branes. The configurations of these D-branes in theories with finite $N_f$ and large $N_c$ are determined by finding action-minimizing configurations of the branes on a fixed background geometry. On the other hand, in order to have $N_f$ of order $N_c$ in a large $N_c$ theory, we need a large number of these flavor branes, and these will back-react on the spacetime geometry itself. For $N_f \sim N_c$, there are as many degrees of freedom in the flavor fields as there are in the color fields (gauge fields and adjoints), so it is natural to expect that the back-reaction will be so significant that in the final description the flavor branes themselves will be completely replaced by a modified geometry with fluxes (in the same way that the branes whose low-energy excitations give rise to the adjoint degrees of freedom do not appear explicitly in the gravity dual description of the field theory).

There has been significant progress in understanding the back-reaction of flavor branes, with some fully-back reacted analytic solutions available (for a review see \cite{Nunez:2010sf}), but so far, there has not been enough progress to fully explore the phase structure of a QCD-like theory with finite $N_f/N_c$. In particular, as far as we are aware, color superconductivity phases have not been identified previously in holographic field theories.\footnote{However, see \cite{Chen:2009kx} for a possible manifestation of the related color-flavor locking phase in a holographic system.}

\subsubsection*{Quark matter from the bottom up}

In this paper, we aim to come up with a holographic system describing a confining gauge theory that does exhibit a quark-matter phase with color superconductivity at large chemical potential. However, motivated by  recent condensed matter applications of gauge/gravity duality (see, for example \cite{Gubser:2008px}), we will avoid many of the technical challenges described above by taking what is known as a ``bottom up'' approach. Rather than working in a specific string theoretical model which takes into account the back-reaction of flavor branes, we will make an ansatz for the ingredients necessary for such a model to describe the relevant physics. We study the simplest possible gravitational theory with this minimal set of features, with the hope that it captures the qualitative physics of interest. We will indeed find that even this simple theory exhibits many of the expected features.

\subsubsection*{Ingredients}

We wish to construct a gravitational theory to provide a holographic description of a four-dimensional confining gauge theory on Minkowski space with $N_f \sim N_c$ flavors. On the gravity side, the Minkowski space will appear as the fixed boundary geometry of our spacetime, but we must have at least one extra dimension corresponding to the energy scale in the field theory. Since the field theory has a scale (the QCD or confinement scale), the asymptotic behavior of the solution must exhibit an additional scale relative to the asymptotically AdS geometries that appear in gravity duals of conformal field theories. In the simplest examples of gravity duals for confining gauge theories, this scale is provided by the size of an additional circular direction in the geometry.\footnote{There are other possibilities here, as we mention briefly in the discussion section.} Thus, we will work with a gravitational system in six dimensions whose boundary geometry is $R^{3,1} \times S^1$. We will assume that the asymptotic geometry is locally Anti-de-Sitter space, so the confining gauge theory we consider arises from a five-dimensional conformal field theory compactified on a circle. When we study the theory at finite temperature, there will be an additional circle in the asymptotic (Euclidean) geometry, the Euclidean time direction whose period is $1/T$.

The gauge theories we are interested in have at least one other conserved current, corresponding to baryon (or quark) number. By the usual AdS/CFT dictionary, this operator corresponds on the gravity side to a $U(1)$ gauge field in the bulk. The asymptotic value of the time component for this gauge field corresponds to the chemical potential in our theory, while the asymptotic value of the radial electric flux corresponds to the baryon charge density in the field theory. For a given chemical potential, the minimum action solution will have some specific value for the flux, allowing us to relate density and chemical potential.

The color superconductivity phases believed to exist at large density in QCD and related theories are usually characterized by  condensates of the form $\langle\psi \psi \rangle$, bilinear in the quark fields $\psi$, which spontaneously break the $U(N)$ gauge symmetry, and the  $U(1)_B$ global symmetry. Naively, we would want to model such operators by a bulk charged scalar field corresponding to the condensate. However, bulk fields always correspond to gauge-invariant operators, while by definition the $\psi \psi$ bilinears which break the gauge symmetry are not gauge-invariant (in fact, there is no way to make a singlet from two fundamental fields, except in the case of $SU(2)$). Additionally, the simplest gauge-invariant operators charged under $U(1)_B$ involve $N$ $\psi$ fields and have dimension of order $N$, thus our holographic dual theory should have {\it no} light scalar fields charged under the $U(1)_{B}$ gauge field.

The correct way to understand the condensation of the $\psi \psi$ bilinears is as an example of spontaneously broken gauge symmetry (as in the Higgs mechanism), rather than as a phase transition characterized by some gauge-invariant order parameter. Nevertheless, the transition to color superconductivity {\it can} be characterized by the discontinuous behavior of gauge-invariant operators, which are of the form $\psi \psi (\psi \psi)^\dagger$. Such operators are gauge invariant and neutral under the $U(1)_B$, and therefore should correspond to an uncharged scalar field in the bulk with dimension of order 1.\footnote{As emphasized to us by Andreas Karch, a gauge invariant operator of the form ${\cal O}_4 = \psi \psi (\psi \psi)^\dagger$ can be written as a sum of terms $\cal O_\alpha \cal O_\alpha$ where each $O_\alpha \sim (\psi^\dagger \psi)_\alpha$ is gauge invariant (and $\alpha$ represents flavor/Lorentz indices). Thus, ${\cal O}_4$ is something like a double-trace operator. In a large $N$ theory, factorization of correlators implies that the expectation value of ${\cal O}_4$ can be calculated classically from the ${\cal O}_\alpha$ expectation values (up to $1/N$ corrections). Thus, discontinuous behavior of ${\cal O}_4$ should be directly related to discontinuous behavior in the simpler gauge-invariant operators $O_\alpha$ (which also have no baryon charge), so it may be more appropriate to think of the scalar field in our model as being dual to one of these simpler operators.}

Combining everything so far, we want to study gravity in six dimensions with negative cosmological constant and boundary geometry $R^{3,1} \times S^1$ with a $U(1)$ gauge field and a neutral scalar field. The simplest action for this system is\footnote{Since we will also consider the case of a charged scalar field, we have written the action using standard normalizations for a complex scalar, but we will take the scalar to be real in the uncharged case.}
\be
\int d^6 x \sqrt{-g} \left\{ {\cal R} + {20 \over L^2} - {1 \over 4} F^2 - |\partial_\mu \psi|^2 - m^2 |\psi|^2 \right\},
\label{action}
\ee
where we include one tunable parameter, the mass $m$ of the scalar field, which determines the dimension of the corresponding operator in the dual field theory. More generally, we could consider other potentials for the scalar field, or a more complicated action (e.g. with a Chern-Simons term or of Born-Infeld type) for the gauge field, but we restrict here to this simplest possible model.\footnote{For another approach to modeling the QCD phase diagram by an effective holographic approach, see for example \cite{Gursoy:2010fj,DeWolfe:2010he}.}

\subsubsection*{Results}

\begin{figure}
\begin{center}
\includegraphics[width=0.7\textwidth]{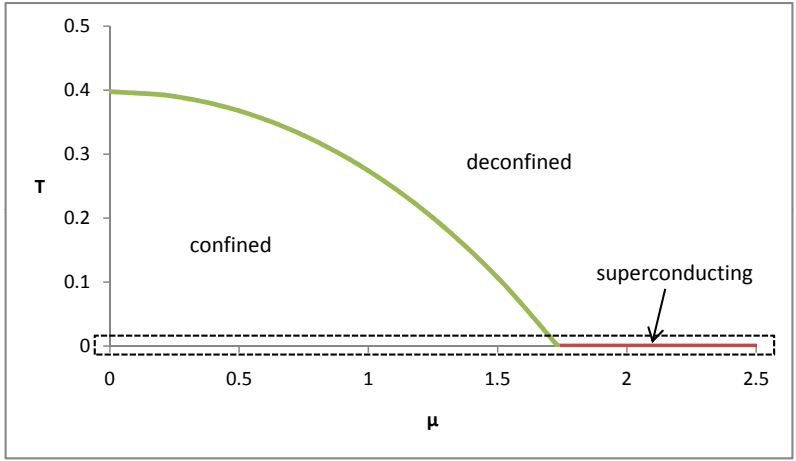}
\end{center}
\caption{\label{neutralphase1} Phase diagram of our model gauge theory with $m^2 = -6$, $R = 2/5$. Region in dashed box is expanded in next figure.}
\end{figure}

\begin{figure}
\begin{center}
\includegraphics[width=0.7\textwidth]{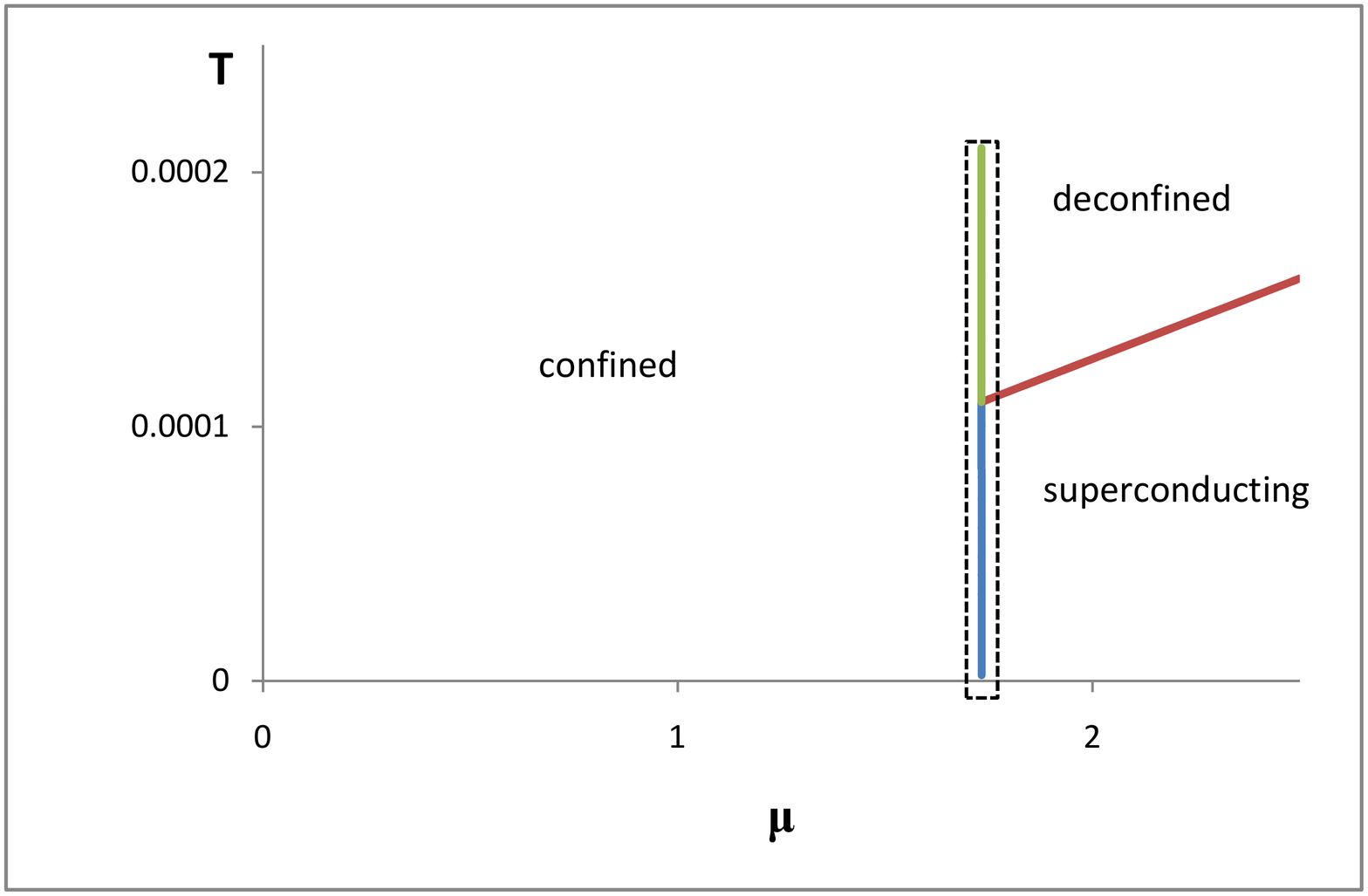}
\end{center}
\caption{\label{neutralphase2} Phase diagram of our model gauge theory with $m^2 = -6$, $R = 2/5$. Region in dashed box is expanded in next figure.}
\end{figure}

\begin{figure}
\begin{center}
\includegraphics[width=0.7\textwidth]{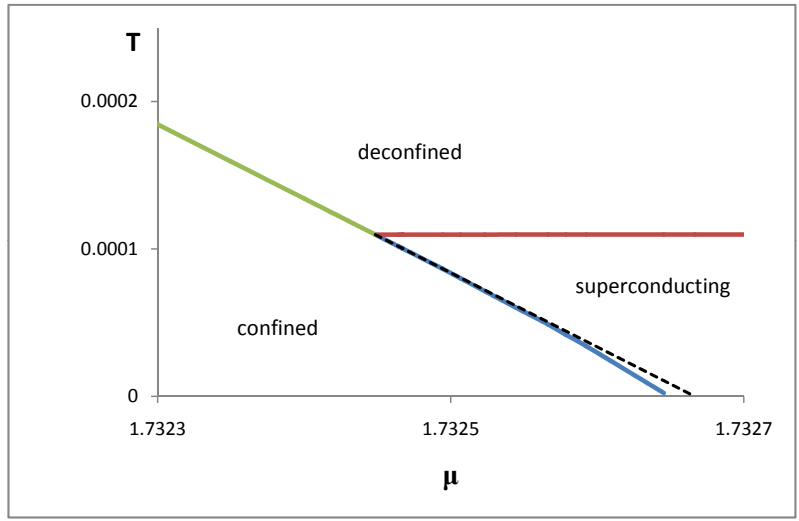}
\end{center}
\caption{\label{neutralphase3} Phase diagram of our model gauge theory with $m^2 = -6$, $R = 2/5$. The dashed curve represents the phase boundary in theory without a scalar field.}
\end{figure}

Starting with the model (\ref{action}), we have explored the phase structure by minimizing the gravitational action for specific values of temperature (corresponding to the asymptotic size of the Euclidean $S^1$ direction) and chemical potential (corresponding to the asymptotic value of $A_0$). Our results for the phase diagrams are shown in figures \ref{neutralphase1},\ref{neutralphase2},\ref{neutralphase3}. For small $\mu$, we find a confined phase at low-temperature and a deconfined phase at high temperature, with the scalar field uncondensed in each case. However, increasing $\mu$ at zero temperature, we find (setting $L_{AdS}=1$) for $-\frac{25}{4} \le m^2 \le -5$ a transition to a phase with nonzero scalar condensate (on a geometry with horizon) and finite homogeneous quark density, as expected for a color superconductivity phase. Increasing the temperature from zero, we find a transition back to the deconfined phase at a remarkably low temperature; for example, at $m^2 = -6$, the critical temperature at which superconductivity disappears is
\[
T /\mu \sim .00006333 \;.
\]

The tendency for the scalar field to condense at low temperatures for the range of masses above can be understood in a simple way, as explained for example in \cite{Hartnoll:2008kx, Iqbal:2010eh}. In $d+1$-dimensional anti-de Sitter space with anti-de Sitter radius $L$, the minimum mass for a scalar field to avoid instability is $m^2_{BF} = -d^2/(4L^2)$. The minimum action solution for large chemical potential in the absence of any scalar field is a planar Reissner-Nordstrom black hole solution with one of the isometry directions periodically identified. In the limit of zero temperature, the near horizon region of this black hole has geometry $AdS^2 \times R^4$, with the radius of the $AdS^2$ equal to $L_2 = L/\sqrt{20}$. Thus, in the near-horizon region, there will be an instability toward condensation of the scalar field if $m^2 < - 1/(4 L_2^2) = -5/L^2$. We thus have a range (setting $L=1$) of $-25/4 \le m^2 \le -5$ for which the scalar field tends to condense in the near-horizon region but is stable in the asymptotic region. Numerical simulations verify that we indeed have scalar field condensation for precisely this range of masses.

While there is no guarantee that the gravitational system we study has a legitimate field theory dual, ``top-down'' gravitational systems corresponding to fully consistent field theories must have the same basic elements (usually with additional fields and a more complicated Lagrangian). The fact that the expected physics emerges even in our stripped-down version suggests that quark-matter phases will be found also in the complete models, once back-reaction effects are under control. Optimistically, qualitative features that we find in the bottom-up model (such as the extremely low transition temperature between superconducting and deconfined phases) may be present also in more complete holographic theories. In this case, our simple model may provide novel qualitative insights into fully consistent QCD-like theories.

\subsubsection*{Charged scalar}

While less relevant to color superconductivity, it is also interesting to explore the physics of our model when we make the scalar field charged under the gauge field. In this case, the scalar field corresponds to a gauge-invariant operator in the field theory that is charged under the $U(1)$ associated with $A$, and the kinetic term for the scalar field is modified in the usual way as $\partial_\mu \psi \to \partial_\mu \psi - i q A_\mu \psi$. As we have argued above, this symmetry cannot be $U(1)_B$, but could be another flavor symmetry, such as isospin in a model with two or more flavors. The flavor superconductivity associated with meson condensation was studied previously in the holographic context (with finite $N_{f}$), for example in \cite{Ammon:2008fc,Basu:2008bh, Ammon:2009fe}. Our results are qualitatively similar to the ones obtained in those studies, and we leave more detailed comparison for future work.

In section 5 below, we determine the phase diagram for various values of $q$ and $m$. The same system was studied for the 2+1 dimensional case in \cite{Horowitz:2010jq} and originally in \cite{Nishioka:2009zj} for the case of large $q$. The application there was to holographic insulator/superconductor systems, but the intriguing resemblance of the phase diagrams in those papers to QCD phase diagrams partially motivated the present study.

\section{Basic setup}

In this paper, we consider holographic field theories with a conserved current $J^\mu$, assumed to be a baryon current (or isospin current when we consider charged scalar fields) and some gauge-invariant operator ${\cal O}$ whose condensation indicates the onset of (color or flavor) superconductivity. We would like to explore the phase structure of the theory for finite temperature $T$ and chemical potential $\mu$; that is, we would like to find the phase that minimizes the Gibbs free energy density $g = e - Ts - \mu \rho$, where $e,s,$ and $\rho$ are the energy density, entropy density, and charge density in the field theory. We can also ask about the values of $e$, $s$, $\rho$, and $\langle {\cal O} \rangle$ as a function of temperature and chemical potential.

As discussed in the introduction, our holographic theories are defined by a dual gravitational background which involves a metric, $U(1)$ gauge field, and scalar field, with a simple action
\be
\int d^6 x \sqrt{-g} \left\{ {\cal R} + {20 \over L^2} - {1 \over 4} F^2 - |\partial_\mu \psi|^2 - m^2 |\psi|^2 \right\} \;.
\label{action2}
\ee
We choose coordinates $(t,x,y,z)$ for the non-compact field theory directions, $w$ for the compact field theory direction, and $r$ for the radial direction. We take boundary conditions for which the asymptotic (large $r$) behavior of the metric is
\beas
ds^2 &\to& \left( {r \over L} \right)^2\, \left( -dt^2 +  dx^2+ dy^2 +dz^2 + dw^2\right)  + \left( {L \over r} \right)^2 dr^2 \;,
\label{asy}
\eeas
where $w$ is taken to be periodic with period $R$. To study the theory at finite temperature, we take the period of $\tau = i t$ in the Euclidean solution to be $1/T$.

The equations of motion constrain the gauge field to behave asymptotically as
\[
A_\nu = a_\nu - {j_\nu \over 3 r^3} + \dots \; .
\]
Since $A_\nu$ is assumed to be the field corresponding to the conserved baryon current operator $J^\nu$, in the field theory, the usual AdS/CFT dictionary tells us that $a_\nu$ is interpreted as the coefficient of the $J^\nu$ in the Lagrangian (i.e. an external source for the baryon current) while $j_\mu$ is interpreted as the expectation value of baryon current for the state corresponding to the particular solution we are looking at. To study the theory at finite chemical potential $\mu$ without any external source for the spatial components of the baryon current, we want to take
\[
a_\nu = (\mu,0,0,0) \; .
\]

The scalar field equations of motion imply that asymptotically
\be
\label{psifall}
\psi = {\psi_1 \over r^{\lambda_-}} + {\psi_2 \over r^{\lambda_+}} + \cdots \;,
\ee
where
\[
\lambda_\mp = {1 \over 2} (d \mp \sqrt{d^2 + 4 m^2}) \; .
\]
The holographic field theories we consider are defined by assuming $\psi_1 = 0$. In this case, $\lambda_+$ gives the dimension of the operator dual to $\psi$.\footnote{For a certain range scalar field masses in the range $-d^2/4 \le m^2 \le -d^2/4 + 1$, it is also consistent to define a theory by fixing $\psi_2 = 0$. In this case, the dimension of the dual operator is $\lambda_-$. We consider this case briefly in section 4.2.} In this case, $\psi_2$ (which will be different for solutions corresponding to different states of the field theory) gives us the expectation value of the operator ${\cal O}$ in the field theory.

By the AdS/CFT correspondence, the field theory free energy corresponds to the Euclidean action of the solution. Thus, to investigate the field theory state which minimizes free-energy for given $T$ and $\mu$, we need to find the gravitational solution with boundary conditions given above which minimizes the Euclidean action. Note that we only consider solutions with translation invariance in $t,x,y,z,$ and $w$. It would be interesting to investigate the possibility of inhomogeneous phases (or at least the stability of our solutions to inhomogeneous perturbations) but we leave this as a question for future work.

\subsection*{Calculating the action}

In order to obtain finite results when calculating the gravitational action for a solution, it is important to include boundary contributions to the action. In terms of the Lorentzian metric, gauge field and scalar, the fully regulated expression that we require is \cite{Gubser:2008px}
\beas
S &=& \lim_{r_M \to \infty} \left[ - \int_{r < r_M} d^{d+1}x \sqrt{-g} \left\{ {\cal R} + {d(d-1) \over L^2} - {1 \over 4} F^2 - |D_\mu \psi|^2 - m^2 |\psi|^2 \right\} \right.\cr
&& \left.+ \int_{r=r_M} d^d x \sqrt{-\gamma} \left\{ -2 K + {2(d-1) \over L}  - {1 \over L} \lambda_- |\psi|^2 \right\} \right] \;,
\eeas
where
\[
\lambda_- = {d \over 2} - {1 \over 2} \sqrt{d^2 + 4 m^2} \;.
\]
Here, $\gamma$ is the metric induced on the boundary surface $r = r_M$, and $K$ is defined as
\[
K = \gamma^{\mu \nu} \nabla_\mu n_\nu \;,
\]
where $n^\mu$ is the outward unit normal vector at $r=r_M$. The scalar counterterm here is the appropriate one assuming that our boundary condition is to fix the coefficient of the leading term in the large $r$ expansion of $\psi$. Since we are setting this term to zero, it turns out that the counterterm vanishes in the $r_M \to \infty$ limit.

For all cases we consider, the metric takes the form
\be
\label{genform}
ds^2 = {r^2 \over L^2} dx_i^2  + g_{00}(r) dt^2 + g_{rr}(r) dr^2 + g_{ww}(r) dw^2 \;.
\ee
Assuming the Einstein equations are satisfied, we can show (by subtracting a term proportional to the $xx$ component of the equation of motion) that the integrand in the first term may be written as a total derivative with respect to $r$
\[
 - \sqrt{-g} \left\{ {\cal R} + {d(d-1) \over L^2} - {1 \over 4} F^2 - |D_\mu \psi|^2 - m^2 |\psi|^2 \right\} = \partial_r \left({2 \over r g_{rr}} \sqrt{-g} \right) \;.
\]
Using
\[
n_\mu = (0,\dots, 0 , \sqrt{g_{rr}}) \;,
\]
we have
\beas
K &=& \gamma^{\mu \nu} \nabla_\mu n_\nu \cr
&=& \gamma^{\mu \nu} \left\{ - \Gamma^r_{\mu \nu} n_r \right\} \cr
&=& \gamma^{\mu \nu} \left\{ {1 \over 2} g^{rr} {\partial g_{\mu \nu} \over \partial r} \sqrt{g_{rr}} \right\} \cr
&=& {1 \over 2 \sqrt{g_{rr}}} \gamma^{\mu \nu} {\partial \gamma_{\mu \nu} \over \partial r} \cr
&=& {1 \over \sqrt{g_{rr}}} {\partial \ln(\sqrt{-\gamma}) \over \partial r} \cr
\eeas
so that
\[
\sqrt{-\gamma}(-2 K) = - {2 \over \sqrt{g_{rr}}}{\partial \sqrt{-\gamma} \over \partial r} \; .
\]
Our final expression for the action density is
\be
\label{acsimp}
S/V_d = \left. {2 \over r g_{rr}} \sqrt{-g} \right|^{r_M}_{r_0} + \left\{- {2 \over \sqrt{g_{rr}}}{\partial \sqrt{-\gamma} \over \partial r} + {2(d-1) \over L} \sqrt{-\gamma} \right\}_{r=r_M} \;.
\ee

\subsection*{Action in terms of asymptotic fields}

It is convenient to rewrite the expression (\ref{acsimp}), in terms of the asymptotic expansion of the fields. For the ansatz (\ref{genform}), and the boundary conditions appropriate to our case, we find
\beas
g_{tt} &=& -r^2 + {g_{tt}^{(3)} \over r^3} + \dots \;, \cr
g_{rr} &=& {1 \over r^2} + {g_{rr}^{(7)} \over r^7} + \dots \;, \cr
g_{ww} &=& r^2 + {g_{ww}^{(3)} \over r^3} + \dots \;, \cr
\psi &=& {\psi^{(3)} \over r^3} + \dots \;, \cr
\phi &=& \mu - {\rho \over 3 r^3} + \dots \;.
\eeas
Inserting these expansions into our expression above for the action we find that (assuming the term at $r=r_0$ vanishes)
\[
S = 5 g_{ww}^{(3)} + 4 g_{rr}^{(7)} - 5 g_{tt}^{(3)} \;.
\]
However, using the equations of motion, we find that $g_{ww}^{(3)} + g_{rr}^{(7)} -  g_{tt}^{(3)}=0$, so we can simplify to:
\be
\label{actcalc}
S = -g_{rr}^{(7)} \; .
\ee
Numerically, it can be a bit tricky to read off $g_{rr}^{(7)}$ because there is also a $1/r^8$ term in the expansion of $g_{rr}$. But using the equations of motion, we can find
\[
g_{rr}^{(8)} = {3 \over 4} (7 + m^2) (\psi^{(3)})^2 \;.
\]
From this, it follows that the combination
\[
-r^7 g_{rr}(r) + r^5 - {3 \over 4} (7+m^2) r^5 \psi^2(r)
\]
behaves like
\[
-g_{rr}^{(7)} + {\cal O} (1/ r^3) \; .
\]
So, we can numerically evaluate the action by taking
\[
S \approx -r_*^7 g_{rr}(r_*) + r_*^5 - {3 \over 4} (7+m^2) r_*^5 \psi^2(r_*) \;,
\]
where $r_*$ is taken to be large but not too close to the cutoff value.

\section{Review: $\psi=0$ solutions}

We begin by considering the solutions for which the scalar field is set to zero.

\subsection{AdS Soliton solution}

At zero temperature and chemical potential, the simplest solution with our boundary conditions is pure AdS with periodically identified $w$. However, assuming antiperiodic boundary conditions for any fermions around the $w$ circle, there is another solution with lower action. This is the AdS soliton \cite{Horowitz:1998ha}, described by the metric (setting $L=1$)
\begin{equation}
ds^2 = r^2\, \left( -dt^2 +  dx^2+ dy^2 +dz^2 + f(r) \,  dw^2\right)  +  {dr^2 \over r^2 f(r)} \;,
\label{confine}
\end{equation}
where
\begin{equation}
\label{fconf}
f(r) = 1 - {r_0^5 \over r^5} \; .
\end{equation}
As long as we choose the period $2 \pi R$ for $w$ such that
\begin{equation}
r_0 =  \frac{2}{ 5 \, R}
\end{equation}
the solution smoothly caps off at $r=r_0$. This IR end of the spacetime corresponds in the field theory to the fact that we have a confined phase with a mass gap. The fluctuation spectrum about this solution corresponds to a discrete spectrum of glueball states in the field theory.

Starting from this solution, we can obtain a solution valid for any temperature and chemical potential, by periodically identifying the Euclidean time direction and setting $A_0 = \mu$ everywhere. Using (\ref{actcalc}) we find that the action for this solution is
\[
S_{sol} = - r_0^5 = - \left( {2 \over 5 R} \right)^5 \; .
\]
The negative value indicates that this solution is preferred over the pure AdS solution with action zero.

\subsection{Reissner-Nordstrom Black Hole Solution}

For sufficiently large temperature and/or chemical potential, the AdS soliton is no longer the $\psi=0$ solution with minimum action. The preferred solution is the planar Reissner-Nordstrom black hole, with metric
\begin{equation}
ds^2 = r^2\, \left( -dt^2 f(r) +  dx^2+ dy^2 +dz^2 + dw^2\right)  +  {dr^2 \over r^2 f(r)} \; ,
\label{BH}
\end{equation}
where
\be
\label{fBH}
f(r) = 1 - \left(1 + {3 \mu^2\over 8r_+^2} \right){r_+^5 \over r^5} + {3 \mu^2 r_+^6 \over 8 r^8} \; ,
\ee
the scalar potential is
\[
\phi(r) = \mu \left( 1 - {r_+^3 \over r^3} \right) \;,
\]
and $w$ is periodically identified as before.

\begin{figure}
\begin{center}
\includegraphics[width=0.7\textwidth]{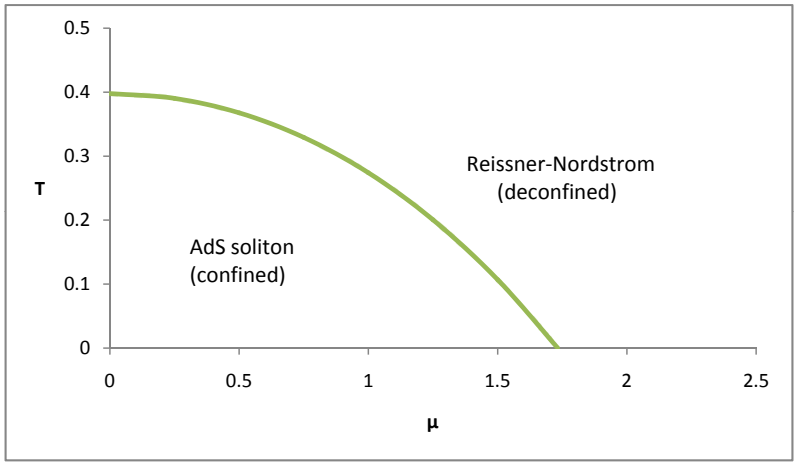}
\end{center}
\caption{\label{neutralphase0} Phase diagram without scalar field, in units where $R=2/5$.}
\end{figure}

This solution has a horizon at $r=r_+$. The temperature of the solution (determined as the inverse period of the Euclidean time for which the Euclidean solution is smooth) is given in terms of $r_+$ by
\be
\label{givesT}
T = {1 \over 4 \pi} \left(5 r_+ - {9 \mu^2 \over 8 r_+} \right) \; .
\ee
From (\ref{actcalc}), we find that the action for this solution is
\[
S_{RN} = - r_+^5 \left(1 + {3 \over 8} {\mu^2 \over r_+^2} \right) \; .
\]
Thus, we find that the black hole solution has lower action than the soliton for
\[
r_+ \left(1 + {3 \over 8} {\mu^2 \over r_+^2} \right)^{1 \over 5} > {2 \over 5 R} \;,
\]
where $r_+$ is determined in terms of $T$ and $\mu$ by (\ref{givesT}). This defines a curve in the $T-\mu$ plane that begins on the $\mu=0$ axis at $T=1 /( 2 \pi R)$ and curves down to the $T=0$ axis at $\mu = 2^{19/10}/(5^{1/2} 3^{4/5} R) \approx 4.3547 /( 2 \pi R)$, as shown in figure \ref{neutralphase0}.

As usual, the existence of a horizon in this solution indicates that the corresponding field theory state is in a deconfined phase \cite{Witten:1998zw}.

In the next sections, we consider solutions with nonzero scalar field. We will find that for large $\mu$ there exist solutions with nonzero scalar field that have lower action than the solutions we have considered, so the phase diagram of figure \ref{neutralphase0} will be modified.

\section{Neutral scalar field: color superconductivity}

In the case of a neutral scalar field, our simple model has no explicit source for the gauge field in the bulk, so homogeneous solutions with a non-trivial static electric field (corresponding to a non-zero baryon number density in the field theory) necessarily have a horizon from which the flux can emerge\footnote{In a more complete model, the source might be provided by some non-perturbative degrees of freedom in the theory, such as the wrapped D-branes that give rise to baryons in the Sakai-Sugimoto model.}.

To look for solutions of this form, we consider the ansatz\footnote{We could have considered a more complicated ansatz, with an extra undetermined function in front of $d w^2$. However, it is plausible that as for the $\psi=0$ solution, the minimum action solution for the case where the $w$ circle does not contract in the bulk is a periodic identification of the solution with non compact $w$ and rotational invariance in the $x,y,z,w$ directions.}
\beas
\label{HR}
ds^2 &=& -g(r)e^{-\chi(r)}dt^2 + {dr^2 \over g(r)} + r^2 (d w^2 + dx^2 + dy^2 + dz^2) \;, \cr
A_t &=& \phi(r) \;, \cr
\psi &=& \psi(r) \;.
\eeas
The scalar and Maxwell's equations that follow from the action (\ref{action2}) are
\begin{equation}
\psi''+\left(\frac{4}{r}-\frac{\chi'}{2}+\frac{g'}{g}\right)\psi'-\frac{m^2}{g}\psi=0\;,
\label{eqfirst}
\end{equation}
\begin{equation}
\phi''+\left(\frac{4}{r}+\frac{\chi'}{2}\right)\phi' =0\;,
\end{equation}
while the Einstein equations are satisfied if
\begin{equation}
\chi'+\frac{r\psi'^2}{2} =0\;,
\end{equation}
\begin{equation}
g'+\left(\frac{3}{r}-\frac{\chi'}{2}\right)g+\frac{re^\chi \phi'^2}{8}+\frac{m^2r\psi^2}{4}-5r=0\;.
\end{equation}

These have two symmetries:
\be
\label{sym1}
\tilde{\psi}(r) = \psi(ar) \;, \qquad \tilde{\phi}(r) = {1 \over a} \phi(ar) \;, \qquad \tilde{\chi}(r) = \chi(ar) \;, \qquad \tilde{g}(r) = {1 \over a^2} g(ar) \;,
\ee
arising from the underlying conformal invariance, and
\be
\label{sym2}
\tilde{\chi} = \chi + \Delta \;, \qquad \tilde{\phi} = e^{-{\Delta \over 2}} \phi \; .
\ee

We would like to find solutions with a horizon at some $r=r_+$. The electric potential must also vanish at the horizon, and we are looking for solutions for which the leading falloff $\psi_1$ in (\ref{psifall}) vanishes for the scalar. Also, multiplying the first equation (\ref{eqfirst}) by $g$ and evaluating at $r=r_+$ fixes $\psi'(r_+)$ in terms of $\psi(r_+)$ and $g'(r_+)$. Altogether, our boundary conditions are
\[
g(r_+) = 0 \;, \qquad \phi(r_+) = 0 \;, \qquad \chi(\infty) = 0 \;, \qquad \psi_1 = 0 \;,
\]
and
\[
\psi'(r_+) = {8 m^2 \psi(r_+) \over 40 r_+ - 2 m^2 r_+^2 \psi^2(r_+)  - r_+ e^{\chi(r_+)} (\phi'(r_+))^2 }\; .
\]
The remaining freedom to choose $r_+$ and $\phi'(r_+)$ leads to a family of solutions with different $T$ and $\mu$. Explicitly, we have
\[
\mu = \phi(\infty) \;, \qquad \qquad T = {1 \over 4 \pi} g'(r_+)e^{- \chi(r_+)/2} \; .
\]
Note that solutions with the same $T/\mu$ are simply related by the scaling symmetry (\ref{sym1}).

\subsection{Numerical evaluation of solutions}

To find solutions in practice, we can make use of the symmetries (\ref{sym1}) to initially set $r_+ = 1$ and $\chi(0)=0$ and solve the equations with boundary conditions
\[
g(1) = 0 \;, \qquad \chi(0) = 0 \;, \qquad \phi(1) = 0 \;, \qquad \phi'(1) = E_0 \;, \qquad \psi(1) = \psi_0 \;,
\]
and
\[
\psi'(1) = {8 m^2 \psi_0 \over 40 - 2 m^2 \psi_0^2  - E_0^2 } \; .
\]
We can integrate the $\phi$ and $\chi$ equations explicitly to obtain
\beas
\chi(r) &=& - \int_0^r d \tilde{r} {1 \over 2} \tilde{r} \left( {\partial \psi \over \partial r} \right)^2 \;, \cr
\phi(r) &=& E_0 \int_1^r  {d \tilde{r} \over \tilde{r}^4} e^{-{1 \over 2} \chi(\tilde{r})} \;,
\eeas
leaving the remaining equations
\beas
\psi'' + ({4 \over r} + {r \psi'^2 \over 4}+ {g' \over g}) \psi' - {m^2 \over g} \psi &=& 0 \;, \cr
g' +  {3 g \over r} + {g r \over 4} \psi'^2 + {E_0^2 \over 8 r^7} + {m^2 r \psi^2 \over 4}-5 r &=& 0 \;.
\eeas

We use $E_0$ as a shooting parameter to enforce $\psi_1 = 0$, and find one solution for each $\psi_0$. From these solutions, we apply the symmetry (\ref{sym2}) with $\Delta = - \chi(\infty)$ to restore $\chi(\infty)=0$ and finally use the symmetry (\ref{sym1}) to scale to the desired temperature or chemical potential.

Using this method, we find that solutions exist for scalar mass in the range $-25/4 \le m^2 \le 5$, which is exactly the range of masses for which the scalar is stable in the asymptotic region but unstable in the near-horizon region.\footnote{Solutions of this form were first found in lower dimensions in \cite{Hartnoll:2008kx}. The zero-temperature limit of such solutions were considered in \cite{Horowitz:2009ij}.} For a given $m^2$ in this range, solutions exist in the region $T/\mu < \gamma(m^2)$, where $\gamma(m^2)$ is a dimensionless number depending on $m^2$ (which we evaluate in the next section). The value of $\gamma(m^2)$ is remarkably small for all $m^2$ in the allowed range. For example, with $m^2= -6$ (not particularly close to the limiting value $m^2 = -5$), we have $\gamma \approx .00006333$. It would be interesting to understand better how this small dimensionless number emerges since the setup has no small parameters. From the bulk point of view it is presumably related to the warping between IR and UV regions of the geometry.\footnote{By considering the alternate quantization mentioned in section 2 and fine-tuning the mass so that the dual operator has the smallest possible dimension consistent with unitarity in the dual field theory, we can obtain $\gamma$ as large as $0.0151$, so even under the most favorable circumstances, the critical $T/\mu$ is quite small.} From the boundary viewpoint, the low critical temperature may be explained by the BKL scaling \cite{Kaplan:2009kr, Jensen:2010ga,  Iqbal:2010eh} near a quantum critical point\footnote{We thank D.T. Son for pointing this out to us.}.

For a given $T$ and $\mu$, we can use (\ref{actcalc}) to evaluate the action for the solution and compare this with the action for the soliton and/or Reissner-Nordstrom solution with the same $T$ and $\mu$. We find that the action for the new solutions is always less than the action for the Reissner-Nordstrom solutions, and is also less than the action for the soliton solutions for chemical potential in a region $\mu > \mu_c(T)$. Thus, the solutions with scalar field represent the equilibrium phase in the region $T/\mu < \gamma, \mu > \mu_c(T)$, as shown in figures \ref{neutralphase1}- \ref{neutralphase3} above.

The transition between the deconfined and superconducting phases is second order, while the transition between confined and superconducting phases is first order. The place where these phase boundaries meet represents a triple point for the phase diagram where the three phases (confined, deconfined, superconducting) can coexist.

\subsection{Critical temperature}

For fixed $m^2$, the value of $\psi(0)$ in the solutions increases from zero at $T/\mu = \gamma$, diverging as $T/\mu \to 0$. Since $\psi$ is small everywhere near $T/\mu = \gamma$, the critical value of $T/\mu$ will be the value where the $\psi$ equation, linearized around the Reissner-Nordstrom background, has a solution with the correct boundary conditions. Thus, we consider the equation
\be
\psi'' + ({4 \over r} + {g' \over g}) \psi' - {m^2 \over g} \psi = 0 \;,
\ee
where (setting $r_+=1$)
\[
g(r) = r^2 - \left(1 + {3 \mu^2\over 8} \right){1 \over r^3} + {3 \mu^2 \over 8 r^6} \; ,
\]
and find the value $\mu = \mu_c$ for which the equation admits a solution with boundary conditions $\psi(1)=1$ (we are free to choose this),  $\psi'(1) = m^2 / g'(1)$ and the right falloff ($\psi_1=0$) at infinity.\footnote{To obtain a very accurate result, we first find a series solution $\psi_{low}$ near $r=1$ with $\psi(1)=1$ (we are free to choose this) and $\psi'(1) = m^2 / g'(1)$ and find a series solution $\psi_{high}$ for large $r$ with the correct fall-off ($\psi_1=0$) at infinity. Starting with $\psi_{low}$ and $\psi'_{low}$ at some $r=r_1$ where the low $r$ series solution is still very accurate, we then numerically integrate up to $r=r_2$ where the large $r$ series is very accurate and then find $\mu$ for which $\psi'_{num}(r_2)/\psi_{num}(r_2) = \psi'_{high}/\psi_{high}$.}

\begin{figure}
\begin{center}
\includegraphics[width=0.7\textwidth]{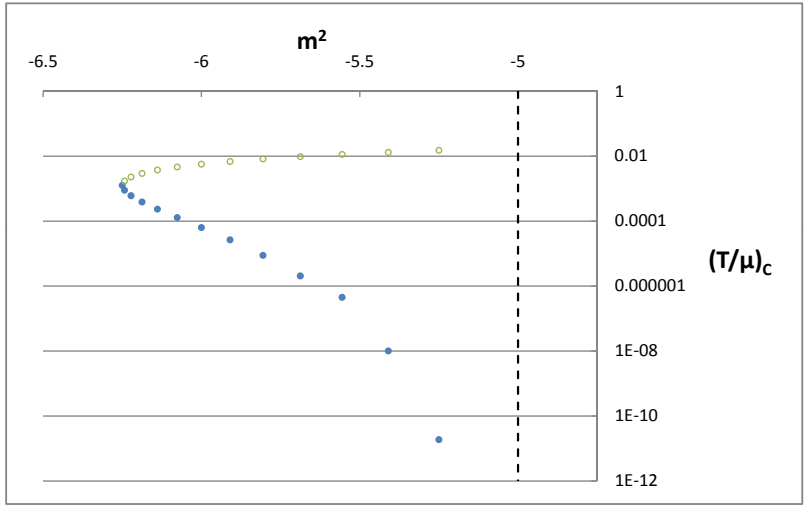}
\end{center}
\caption{\label{tmucvsm} Critical $T/\mu$ vs $m^2$ of neutral scalar (filled circles). Mass is above BF bound asymptotically but below BF bound in near-horizon region of zero-temperature background solution in the range $-6.25 \le m^2 < -5$. Unfilled circles represent critical values in the theory with alternate quantization of the scalar field, possible in the range $-6.25 \le m^2 < -5.25$.}
\end{figure}

The choice $r_+=1$ implies that $T = (5 - 9 \mu^2/8)/(4 \pi)$, so we have $\gamma = (5 - 9 \mu_c^2/8)/(4 \pi \mu_c)$. The results for $\gamma(m^2)$ are plotted in figure \ref{tmucvsm}. For comparison, we also considered the theory defined with the alternate quantization ($\psi_2^\infty=0$) of the bulk scalar field (mentioned in section 2). As we see in figure \ref{tmucvsm}, the critical temperatures are somewhat larger in this case, but still much smaller than 1 relative to $\mu$.


\subsection{Properties of the superconducting phase}

In the superconducting phase, it is interesting to ask how the charge density and free energy behave as a function of chemical potential. Since the solutions (as for the planar RN-black hole solutions) are trivially related to solutions where the $w$ direction is non-compact, and since the underlying theory has a conformal symmetry, physical quantities in this phase (or in the RN phase) behave as $\mu^n F(T/\mu)$ for some non-trivial function $F$ and a power $n$.\footnote{If the solutions instead depended on the circle direction in a non-trivial way, we might have a general function of $R T$ and $R \mu$.} At the critical value of $T/\mu$, we have a second order transition from the RN phase to the phase with scalar, so the free energy and its derivatives, and other physical quantities such as the density, are continuous across the transition. Thus, the relevant function $F$ in these cases will be the same for the two phases across the transition. We find that the function $F$ for either the charge density or the free energy changes very little between the very small value of $T/\mu$ where the the transition occurs and the $T \to 0$ limit. Thus, to a good approximation, we find that the density and free energy behave in the superconducting phase in the same way as for the zero temperature limit of the RN phase. For R=2/5, we have
\[
\rho \approx 0.320 \mu^4 \; ,
\]
while
\[
G \approx -.064 \mu^5 \; .
\]
In both cases, the behavior at large $\mu$ is governed by the underlying 4+1 dimensional conformal field theory.

\section{Charged scalar field: flavor superconductivity}

In this section, we generalize our holographic model to the case where the scalar field is charged under the gauge field in the bulk. As we discussed in the introduction, this implies that the dual field theory includes some low-dimension gauge-invariant operator with charge, so the charge in this case is more naturally thought of as some isospin-type charge (since the smallest gauge-invariant operators carrying baryon charge have dimensions of order $N$).

A significant qualitative difference in this case is that a scalar field condensate acts as a source for the electric field in the bulk, so it is possible to have solutions with no horizon carrying a finite charge density in the field theory. This gives the possibility of a fourth phase in which the scalar field condenses in the soliton background.

To obtain the action for the charged scalar case, we begin with the action (\ref{action2}) and make the replacement $\partial_\mu \psi \to \partial_\mu \psi - i q A_\mu \psi$. The results of the previous section correspond to $q=0$.

\subsection{Low-temperature horizon free solutions with scalar}

Above some critical value of $\mu$, there exist horizon-free geometries with a scalar field condensate. The solutions may be parameterized by the magnitude of the scalar at the IR tip of the geometry, and we will find a single solution for each such value. To determine these geometries, we need to take into account back-reaction on the metric. The most general solution with the desired properties can be described by the ansatz
\bea
ds^2 &=& r^2(e^{A(r)}B(r) d w^2 + dx^2 + dy^2 + dz^2 - e^{C(r)}dt^2) + {dr^2 \over r^2 B(r)}\; , \cr
A_t &=& \phi(r) \;, \cr
\psi &=& \psi(r) \;,
\label{lowT}
\eea
where we demand $A(\infty) = C(\infty) = 0$ and  $B(\infty) = 1$. As for the soliton geometry, we expect that the $w$ circle is contractible in the bulk so that $B(r_0)=0$ for some $r_0$. For the geometry to be smooth at this point, the periodicity of the $w$ direction must be chosen so that
\be
\label{solrad1}
2 \pi R = {4 \pi e^{-A(r_0)/2} \over r_0^2 B'(r_0)} \; .
\ee

Starting from the action (\ref{action2}) with scalar derivatives replaced by covariant derivatives, the scalar and Maxwell equations are:
\begin{equation}
\psi''+\left(\frac{6}{r}+\frac{A'}{2}+\frac{B'}{B}+\frac{C'}{2}\right)\psi'+\frac{1}{r^2B}\left(\frac{e^{-C}(q \phi)^2}{r^2}-m^2\right)\psi=0\;,
\end{equation}
\begin{equation}
\phi''+\left(\frac{4}{r}+\frac{A'}{2}+\frac{B'}{B}-\frac{C'}{2}\right)\phi'-\frac{2\psi^2q^2 \phi}{r^2 B}=0\;.
\end{equation}
Following \cite{Horowitz:2010jq}, we find that the Einstein equations give:
\begin{equation}
A'=\frac{2r^2C''+r^2C'^2+4rC'+4r^2\psi'^2-2e^{-C}\phi'^2}{r(8+rC')}\;,
\end{equation}
\begin{equation}
C''+{1 \over 2} C'^2+\left(\frac{6}{r}+\frac{A'}{2}+\frac{B'}{B}\right)C'-\left(\phi'^2+\frac{2(q\phi)^2\psi^2}{r^2B}\right)\frac{e^{-C}}{r^2}=0\; ,
\end{equation}
\begin{equation}
B'\left(\frac{4}{r}-\frac{C'}{2}\right)+B\left(\psi'^2-{1 \over 2} A'C'+\frac{e^{-C}\phi'^2}{2r^2}+\frac{20}{r^2}\right)+\frac{1}{r^2}\left(\frac{e^{-C}(q\phi)^2\psi^2}{r^2}+m^2\psi^2-20\right)=0\;.
\end{equation}

These equations have two scaling symmetries,
\be
\label{sc1}
\tilde{\psi}(r) = \psi(ar)\;, \qquad \tilde{\phi}(r) = {1 \over a} \phi(ar)\;, \qquad \tilde{A}(r) = A(ar)\;, \qquad \tilde{B}(r) =  B(ar)\;, \qquad \tilde{C}(r) = C(ar)\;,
\ee
and
\be
\label{sc2}
\tilde{C} = C + \Delta \;, \qquad \tilde{\phi} = e^{{\Delta \over 2}} \phi \; .
\ee

\subsubsection*{Numerical evaluation of solutions}

To find solutions, we first use the scaling symmetries to fix $r_0=1$ and $C(r_0)=0$. For each value of $\psi(1)$, we use $\phi(1)$ as a shooting parameter, choosing the value so that $\psi$ has the desired behavior for large $r$. From the solution obtained in this way, we can use (\ref{sc2}) with $\Delta = - C(\infty)$ to obtain the desired boundary condition $C(\infty)=0$ in the rescaled solution. From (\ref{solrad1}), we see that the choice $r_0=1$ corresponds to a periodicity for the $w$ direction equal to
\be
\label{solrad2}
2 \pi R = {4 \pi e^{-A(1)/2} \over B'(1)} \; .
\ee
which will generally be different for solutions corresponding to different values of $\psi(1)$. In order to obtain solutions corresponding to our chosen value $R=2/5$ (such that the action for the soliton solution is -1) we use the scaling (\ref{sc1}), taking $a=B'(1)/5 e^{-A(\infty)/2}$. After all the scalings, we calculate the chemical potential and action (making use of (\ref{actcalc})) as
\[
\mu = \phi(\infty) \;, \qquad \qquad S = [B]_{1 \over r^5} \; .
\]

\begin{figure}
\begin{center}
\includegraphics[width=0.7\textwidth]{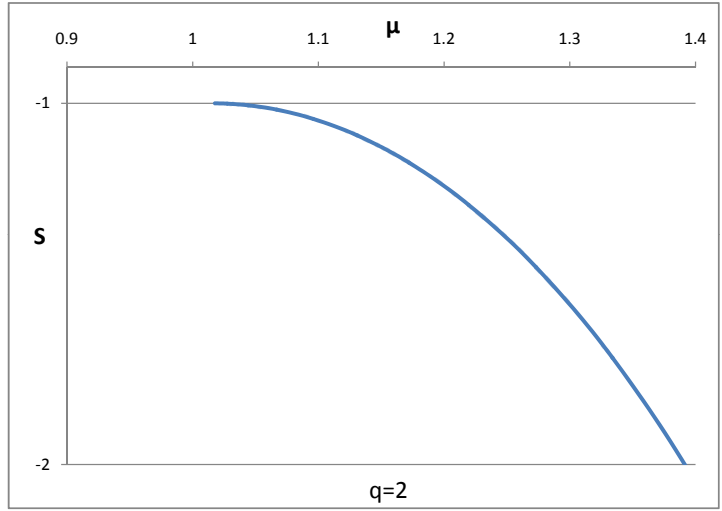}
\end{center}
\caption{\label{sol20} Action vs chemical potential for soliton with scalar solutions, taking $m^2=-6$ and $q=2$.}
\end{figure}

\begin{figure}
\begin{center}
\includegraphics[width=0.7\textwidth]{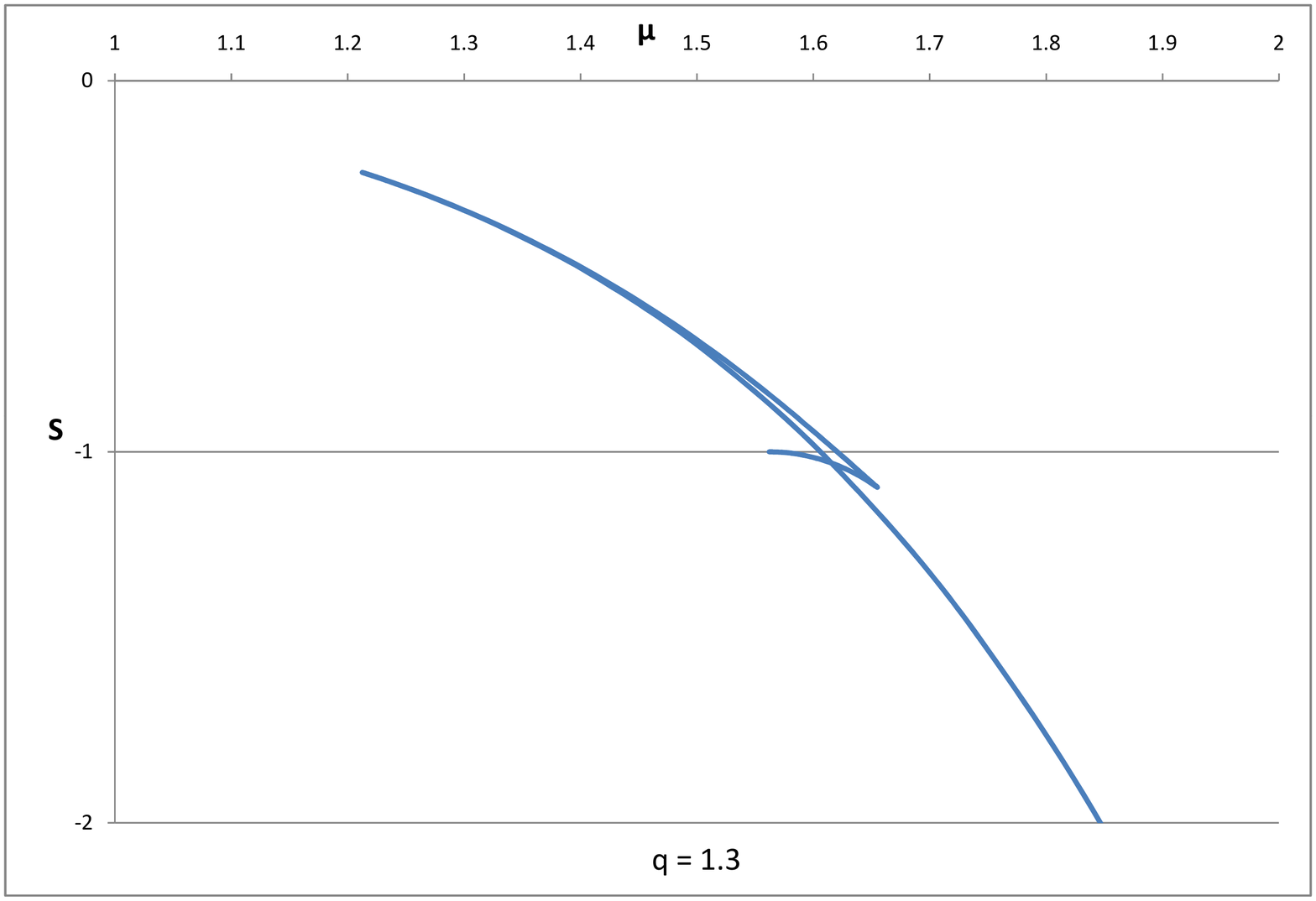}
\end{center}
\caption{\label{sol13} Action vs chemical potential for soliton with scalar solutions, taking $m^2=-6$ and $q=1.3$.}
\end{figure}

\begin{figure}
\begin{center}
\includegraphics[width=0.7\textwidth]{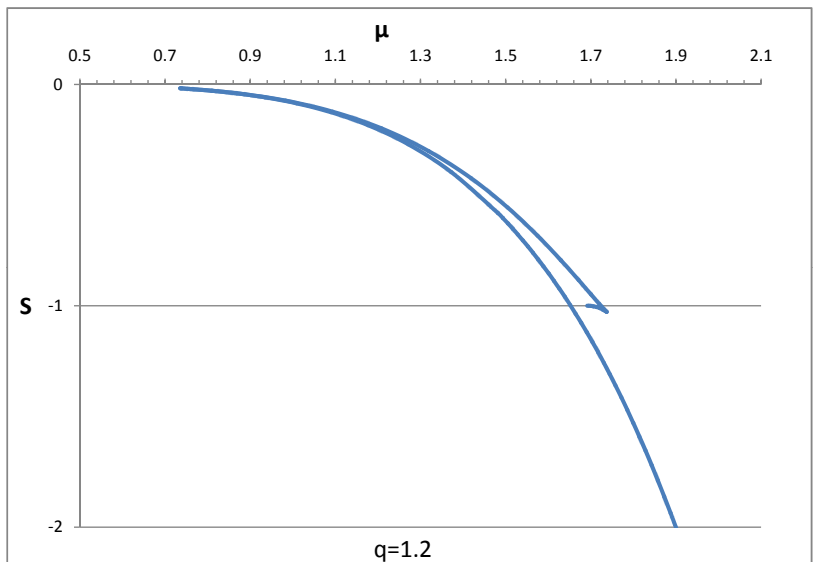}
\end{center}
\caption{\label{sol12} Action vs chemical potential for soliton with scalar solutions, taking $m^2=-6$ and $q=1.2$.}
\end{figure}

The action is plotted against chemical potential for various values of $q$ in figures \ref{sol20}, \ref{sol13}, and \ref{sol12} taking the example of a mass just above the BF bound, $m^2 = -6$.

We find that for large enough values of $q$, the chemical potential increases monotonically and the action decreases monotonically as we increase $\psi(r_0)$. This implies that we have a second order transition to the superconducting phase at a critical value, which can be determined by a linearized analysis (see appendix A) to be  $\mu  \approx 1.0125/q$.

Below $q \approx 1.35$, the chemical potential is no longer monotonic in $\psi(r_0)$. We see that for $q=1.3$, this results in a second order phase transition at $\mu \approx 1.558$, followed by a first order phase transition at $\mu \approx 1.616$ (taking $R=2/5$). For smaller $q$ (e.g. $q=1.2$ in figure \ref{sol12}), we simply have a first order transition to the superconducting phase at a value of chemical potential that is less than the value for the solution with infinitesimal scalar field. All of these results are completely analogous to the lower-dimensional results of \cite{Horowitz:2010jq}.

\subsection{Hairy black hole solutions}

At high temperatures, the $w$ circle is no longer contractible, and we assume that (as for the solutions without scalar field) the solution can be obtained by periodic identification of a solution with boundary $R^{4,1}$ instead of $R^{3,1} \times S^1$. Thus, we take the ansatz
\beas
\label{HR}
ds^2 &=& -g(r)e^{-\chi(r)}dt^2 + {dr^2 \over g(r)} + r^2 (d w^2 + dx^2 + dy^2 + dz^2) \;, \cr
A_t &=& \phi(r) \;, \cr
\psi&=&\psi(r) \;.
\eeas
The scalar and Maxwell's equations are
\begin{equation}
\psi''+\left(\frac{4}{r}-\frac{\chi'}{2}+\frac{g'}{g}\right)\psi'+\frac{1}{g}\left(\frac{e^\chi q^2\phi^2}{g}-m^2\right)\psi=0\;,
\label{eqfirst2}
\end{equation}
\begin{equation}
\phi''+\left(\frac{4}{r}+\frac{\chi'}{2}\right)\phi'-\frac{2q^2\psi^2}{g}\phi=0\;,
\end{equation}
while the Einstein equations are satisfied if
\begin{equation}
\chi'+\frac{r\psi'^2}{2}+\frac{re^\chi q^2\phi^2\psi^2}{2g^2}=0\;,
\end{equation}
\begin{equation}
g'+\left(\frac{3}{r}-\frac{\chi'}{2}\right)g+\frac{re^\chi \phi'^2}{8}+\frac{m^2r\psi^2}{4}-5r=0\;.
\end{equation}
These have two symmetries:
\begin{equation}
\label{sym3}
\tilde{\psi}(r) = \psi(ar) \;, \qquad \tilde{\phi}(r) = {1 \over a} \phi(ar) \;, \qquad \tilde{\chi}(r) = \chi(ar) \;, \qquad \tilde{g}(r) = {1 \over a^2} g(ar) \;,
\end{equation}
and
\begin{equation}
\label{sym4}
\tilde{\chi} = \chi + \Delta \;, \qquad \tilde{\phi} = e^{-{\Delta \over 2}} \phi \; .
\end{equation}
As we did for $q=0$, we would like to find solutions with a horizon at some $r=r_+$. The electric potential must also vanish at the horizon, and we are looking for solutions for which the leading falloff $\psi_1$ in (\ref{psifall}) vanishes for the scalar. Also, multiplying the first equation (\ref{eqfirst2}) by $g$ and evaluating at $r=r_+$ fixes $\psi'(r_+)$ in terms of $\psi(r_+)$ and $g'(r_+)$. Altogether, our boundary conditions are
\[
g(r_+) = 0 \;, \qquad \phi(r_+) = 0 \;, \qquad \chi(\infty) = 0 \;, \qquad \psi_1 = 0 \;,
\]
and
\[
\psi'(r_+) = {8 m^2 \psi(r_+) \over 40 r_+ - 2 m^2 r_+^2 \psi^2(r_+)  - r_+ e^{\chi(r_+)} (\phi'(r_+))^2 }\; .
\]
The remaining freedom to choose $r_+$ and $\phi'(r_+)$ leads to a family of solutions with different $T$ and $\mu$. Explicitly, we have
\[
\mu = \phi(\infty) \;, \qquad \qquad T = {1 \over 4 \pi} g'(r_+)e^{- \chi(r_+)/2} \; .
\]
Solutions with the same $T/\mu$ are simply related by the scaling symmetry (\ref{sym3}).

\subsubsection*{Numerical evaluation of solutions}

To find solutions in practice, we can make use of the symmetries (\ref{sym3}, \ref{sym4}) to initially set $r_+ = 1$ and $\chi(0)=0$ and solve the equations with boundary conditions
\[
g(1) = 0 \;, \qquad \chi(0) = 0 \;, \qquad \phi(1) = 0 \;, \qquad \phi'(1) = E_0 \;, \qquad \psi(1) = \psi_0 \;,
\]
and
\[
\psi'(1) = {8 m^2 \psi_0 \over 40 - 2 m^2 \psi_0^2  - E_0^2 } \; .
\]

We use $E_0$ as a shooting parameter to enforce $\psi_1 = 0$, and find one solution for each $\psi_0$. From these solutions, we apply the symmetry (\ref{sym4}) with $\Delta = - \chi(\infty)$ to restore $\chi(\infty)=0$ and finally use the symmetry (\ref{sym3}) to scale to the desired temperature or chemical potential.

\subsection{Phase diagrams}

At a generic point in the phase diagram, we can have up to four solutions (AdS soliton, planar RN black hole, soliton with scalar, black hole with scalar), or more in cases where there is more than one solution of a given type.

\begin{figure}
\begin{center}
\includegraphics[width=0.7\textwidth]{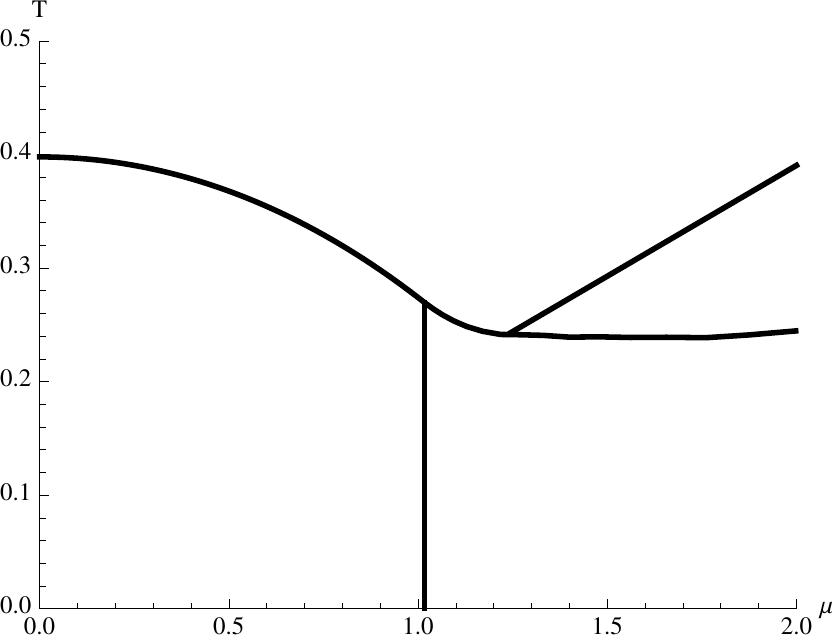}
\end{center}
\caption{\label{phase20} Phase diagram for $m^2=-6$ and $q=2$. Clockwise from the origin, the phases correspond to the AdS soliton (confined), RN black hole, black hole with scalar, and soliton with scalar. }
\end{figure}

\begin{figure}
\begin{center}
\includegraphics[width=0.7\textwidth]{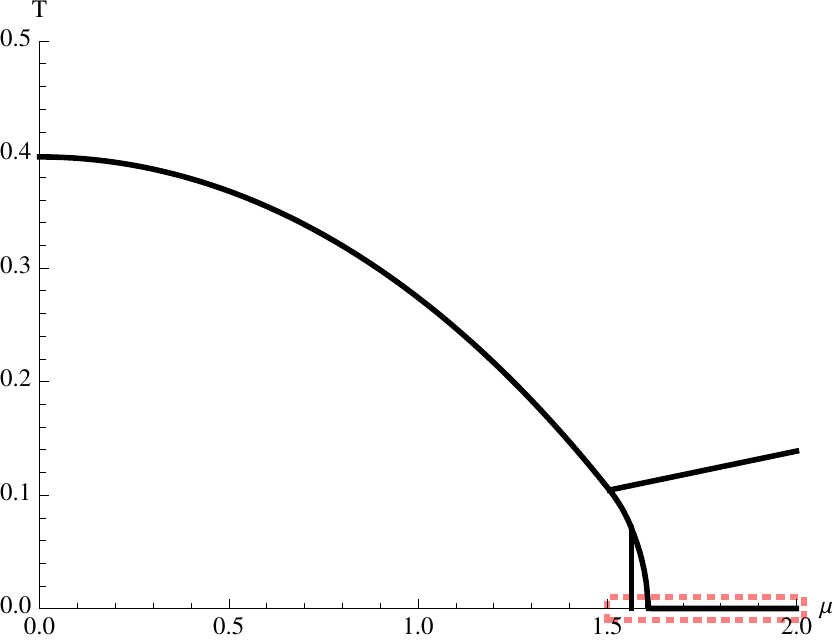}
\end{center}
\caption{\label{phase13a} Phase diagram for $m^2=-6$ and $q=1.3$. Clockwise from the origin, the phases correspond to the AdS soliton (confined), RN black hole, black hole with scalar, and soliton with scalar.}
\end{figure}

\begin{figure}
\begin{center}
\includegraphics[width=0.7\textwidth]{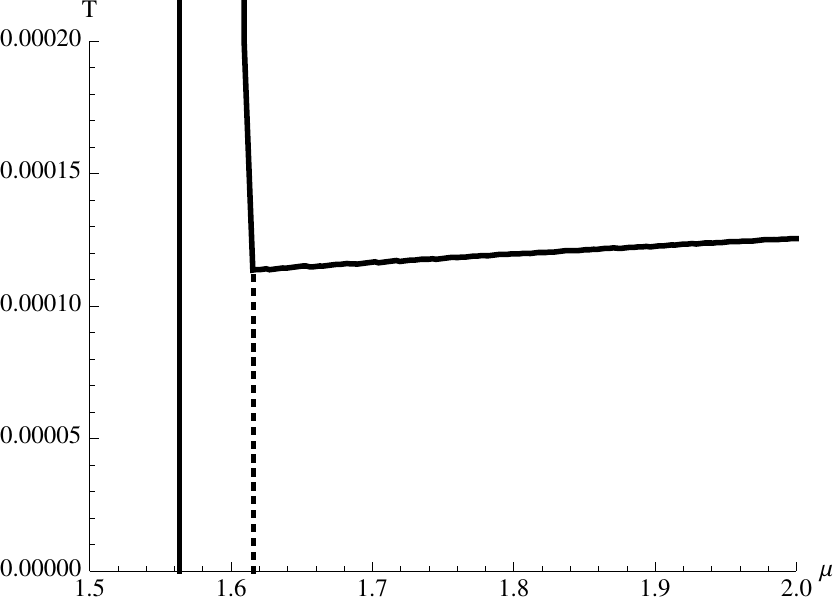}
\end{center}
\caption{\label{phase13b} Small temperature region of phase diagram for $m^2=-6$ and $q=1.3$. Dashed line represents a first order transition within the soliton with scalar phase.}
\end{figure}

To map out the phase diagram, we evaluate the action for the various solutions using the methods of section 2. The equilibrium phase corresponds to the solution with lowest action. The phase diagrams for $q=1.3$ and $q=2$ (in the case $m^2 = -6$) are shown in figures \ref{phase20} and \ref{phase13a}/\ref{phase13b}).

\begin{figure}
\begin{center}
\includegraphics[width=0.7\textwidth]{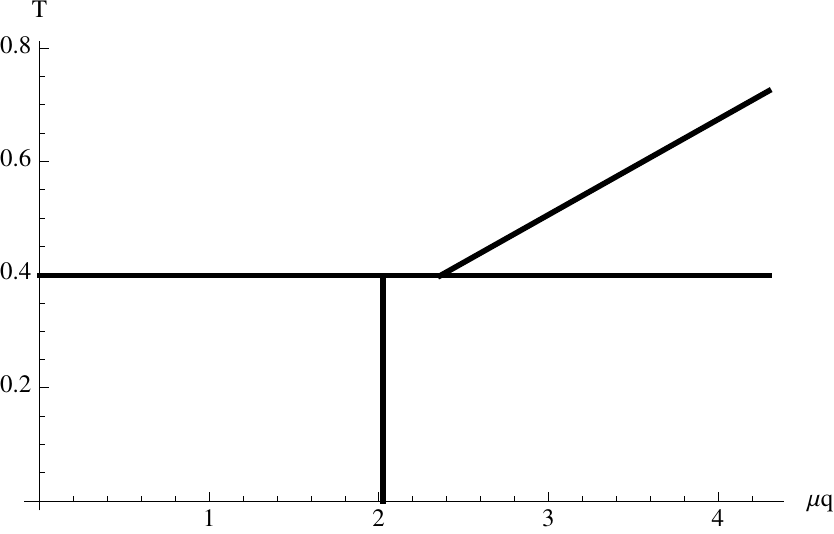}
\end{center}
\caption{\label{phaseq} Phase diagram for large $q$, $m^2 = -6$. }
\end{figure}

For large $q$, the condensation of the scalar field occurs in a region of the phase diagram where the back-reaction is negligible, so the phase diagram may be understood here (for $\mu \sim 1/q$) by treating the gauge field and scalar on a fixed background (the Schwarzschild black hole). The resulting phase diagram is shown in figure (\ref{phaseq}).

\section{Discussion}

In this note, we have investigated the phase structure for a simple class of holographic systems which we have argued have the minimal set of ingredients to holographically describe the phenomenon of color superconductivity. Even in these simple models, we find a rich phase structure with features similar to the conjectured behavior of QCD at finite temperature and baryon chemical potential. It would be useful to verify the thermodynamic stability (and also the stability towards gravitational perturbations) of the phases that we have identified. This could indicate regions of the phase diagram where we have not yet identified the true equilibrium phase for the model, for example since our ansatz might be too symmetric.

We have calculated some of the basic thermodynamic observables, but it would be interesting to investigate more fully the physical properties of the various phases and establish more definitively a connection between the phase we find at large $\mu$ and small temperature and the physics of color superconductivity.

Apart from the $\psi \psi \psi^\dagger \psi^\dagger$ condensate that we can see directly using the ingredients of our model, there are various other features that characterize a color superconductivity phase \cite{Alford:2007xm}. Typically, the breaking of gauge symmetry is accompanied by some breaking of exact or approximate flavor symmetries. Thus, the superconducting phase has a low-energy spectrum characterized by Goldstone bosons or pseudo-Goldstone bosons associated with the broken flavor symmetries, together with massive vector bosons associated with the spontaneously broken gauge symmetry. It would therefore be interesting to analyze the spectrum of fluctuations in our model to compare with these expectations.

A caveat related to looking for features associated with the global flavor symmetries (and their breaking) in our model is that we may not have included enough ingredients in our bottom-up approach for all these features to be present. In simple models where the flavor degrees of freedom are associated with probe branes, there are explicit gauge fields in the bulk dual to the global symmetry current operators. However, in fully back-reacted solutions (appropriate for studying $N_f \sim N_c$), these branes are replaced by a modified geometry with additional fluxes (for an explicit example of such solutions, see \cite{D'Hoker:2007xz}). In these solutions (which we are trying to model in our approach), it is less clear how to identify the global symmetry group from the gravity solution, but presumably it has to do with some detailed properties of the geometry. Thus, it is possible that the Goldstone modes associated with broken flavor symmetries correspond to fluctuations in some fields (e.g. form-fields) that we have not included.

The color superconducting condensate also breaks the global baryon number symmetry, so there should be an associated Goldstone boson related to the phase of the condensate, and associated superfluidity phenomena. In other holographic models with superfluidity, the condensate is dual to a charged scalar field in the bulk and the Goldstone mode is related to fluctuations in the phase of this field. However, as we mentioned in the introduction, the baryon operator has dimension of order $N$, so we do not expect a light charged scalar field in the bulk. In a more complete top-down model, the baryon operator may be related to some non-perturbative degrees of freedom (such as D-branes) in the bulk, and it may be necessary to have a model with these degrees of freedom included in order to directly see the Goldstone mode from the bulk physics. Related to these observations, it may be interesting to probe our model with D-branes (put in by hand), in order to make the relation to microscopic physics more manifest, and to help gain a better understanding of the phenomenological parameters of our model.

There are a number of variants on the model that would be interesting to study. First, the breaking of scale-invariance, implemented in our model by the varying circle direction in the bulk, could be achieved in other ways, replacing $g_{ww}$ with a more general scalar field, as in the model of \cite{DeWolfe:2010he}. In the setup of that paper, the transition between confined and deconfined phases was found to exhibit crossover behavior at small chemical potential, a feature expected in the real QCD phase diagram and expected generally for massive quarks with sufficiently large $N_f/N_c$. It would be interesting to look for an even more realistic holographic model by incorporating features of the model we have studied here and the model of \cite{DeWolfe:2010he}.

It would also be interesting to look at the effects of a Chern-Simons term for the bulk gauge field. In \cite{Nakamura:2009tf} and \cite{Ooguri:2010xs}, it was shown that such a term (with sufficiently large coefficient) gives rise to an instability toward inhomogeneous phases, perhaps associated with the chiral density wave phase believed to exist at large density in QCD with $N_c \ll N_f$ \cite{Deryagin:1992rw,Shuster:1999tn}. It is interesting to investigate the interplay between these inhomogeneous instabilities and the superconducting instabilities discussed in the present paper. It would also be interesting to consider more general actions (such as Born-Infeld) for the gauge field, interaction terms for the scalar field in the bulk, or other couplings between the scalar field and gauge field.

Finally, once the technical challenges of writing down fully back-reacted solutions for top-down models of holographic QCD with $N_f ~ N_c$ have been overcome, it will be interesting to see whether the basic features we find here are manifested in the more complete string-theoretic models. If certain features are found to be universal, these might taken as qualitative predictions for the QCD phase diagram, or at least motivate an effort to understand whether these features are also present in the phase diagram of real-world QCD.

\section*{Acknowledgements}
We thank Ofer Aharony, Oren Bergman, Carlos Hoyos, Don Marolf, Andreas Karch and D.T. Son for useful conversations. The work is supported by NSERC Discovery Grants. PB is supported by NSF grants PHY-0970069 and PHY-0855614.
\appendix

\section{Large charge limit}

In this appendix, we analyze the case of large $q$. This is particularly simple, since in this limit, the back-reaction of the scalar and the gauge field on the metric go to zero in the region of the phase diagram where transitions to the superconducting phases occur. Explicitly, we can show that in the limit $q \to \infty$ with $q \mu$ fixed, the gauge field and scalar field decouple from the equations for the metric, but still give rise to a nontrivial phase structure. To investigate this, we need only consider the scalar field and gauge field equations on the fixed background spacetimes corresponding to low temperatures (the soliton geometry) and high temperatures (the Schwarzschild black hole).

\subsection*{Low Temperature}

Starting from the action (\ref{action2}) for the scalar field and gauge field on the soliton background (\ref{confine}), we find that the equations of motion are (setting $L=1$)
\beas
\phi'' + \left( {f' \over f} + {4 \over r} \right) \phi' - {2 q^2 \over r^2 f} \psi^2 \phi &=& 0 \;, \cr
\psi'' + \left( {f' \over f} + {6 \over r} \right) \psi' + { q^2 \over r^4 f} \phi^2 \psi - {m^2 \over r^2 f} \psi &=& 0 \;, \cr
\eeas
where $f$ is defined in (\ref{fconf}).

These equations have two scaling symmetries related to the conformal symmetry of the boundary field theory and to the absence of back-reaction in our large charge limit. Given a solution $(\phi(r),\psi(r),r_0,q,m)$, we can check that the scaling
\[
(\phi(r),\psi(r),r_0,q,m) \to (\beta \phi(\alpha r),\beta \alpha \psi(\alpha r),{r_0 \over \alpha},{q \over \alpha \beta} ,m)
\]
sends solutions to solutions. For our calculations, we will use this to set $r_0 = q = 1$.

Multiplying these equations by $f$ and taking the limit $r \to r_0=1$, we find that regular solutions must obey
\beas
\phi'(1) &=& {2 \psi^2(1) \phi(1) \over 5 } \;, \cr
\psi'(1) &=& {\psi(1) \over 5}\left(m^2 - \phi^2(1)  \right) \;.
\eeas
We have two remaining parameters, $\psi(0)$ and $\phi(0)$. One of these can be fixed by demanding that the ``non-normalizible'' mode of $\psi$ vanishes at infinity, while different values of the remaining parameter correspond to different values of $\mu$.

Employing numerics, we find that for a fixed value of $m^2$, there is some critical value of $\mu$ above which solutions with a condensed scalar field exist.

In order to determine the critical value $\mu_c(m^2)$, we use the fact that the field values go to zero as we approach the critical $\mu$ from above. Thus, at the critical $\mu$, the equations above linearized around the background solution $\phi = \mu$ should admit a solution with the correct boundary conditions. The linearized equations decouple from each other, so we need only study the $\psi$ equation. This becomes
\[
\psi'' + \left( {6 r^5 -1  \over r(r^5 - 1)} \right) \psi'  + {r( \mu^2 - m^2 r^2) \over r^5 -1} \psi =0 \; .
\]
We can take $\psi(1)=1$ without loss of generality, so the boundary condition for $\psi'$ becomes
\[
\psi'(1) = {1 \over 5}(m^2 - \mu^2) \;.
\]
Given $m^2$, we now find $\mu^2$ by demanding that the leading asymptotic mode ($\psi_1$) of $\psi$ vanishes. Our results for the critical $\mu$ as a function of $m^2$ are shown in figure \ref{mucprobe}.

\subsection*{High temperature}

The high temperature geometry relevant to the limit of large $q$ with $\mu q$ fixed is the $\mu \to 0$ limit of the Reissner-Nordstrom geometry (\ref{BH}), which gives the planar AdS-Schwarzschild black hole (with one of the spatial directions compactified). This is the relevant background for $T > 1/(2 \pi R)$.

Explicitly, we have
\beas
ds^2 &=& r^2\, \left( -dt^2 f(r) +  dx^2+ dy^2 +dz^2 + dw^2\right)  +  {dr^2 \over r^2 f(r)} \;, \cr
\eeas
where
\[
f(r) = 1 - {r_+^5 \over r^5} \;.
\]
Here, $r_+$ is related to the temperature by
\[
r_+ = {4 \pi T \over 5} \;.
\]

The equations of motion in this background are
\beas
 \psi'' + \left( {f' \over f} + {6 \over r} \right) \psi' + { q^2 \over r^4 f^2} \phi^2 \psi - {m^2 \over r^2 f} \psi &=& 0 \;. \cr
\eeas
The equations have the same scaling symmetry as before, so we can set $r_+=q=1$ for numerics. Here, the choice $r_+=1$ corresponds to $T = 1/(2 \pi R)$, where $R$ is the radius chosen in the previous section by setting $r_0=1$. In this case, the boundary conditions are
\[
\phi(1) = 0 \;, \qquad \qquad \psi'(1) = {m^2 L^2 \psi(1) \over 5} \; .
\]
To determine the physics at other temperatures, we can fix $q$ and $R$ and use the scaling to adjust the temperature.

For any values of parameters, we have a solution
\[
\psi = 0 \;, \qquad \phi(r) = \mu( 1 - {1 \over r^3}) \; .
\]
corresponding to the pure Reissner-Nordstrom background in the probe limit.

\begin{figure}[t]
\begin{center}
\includegraphics[width=0.7\textwidth]{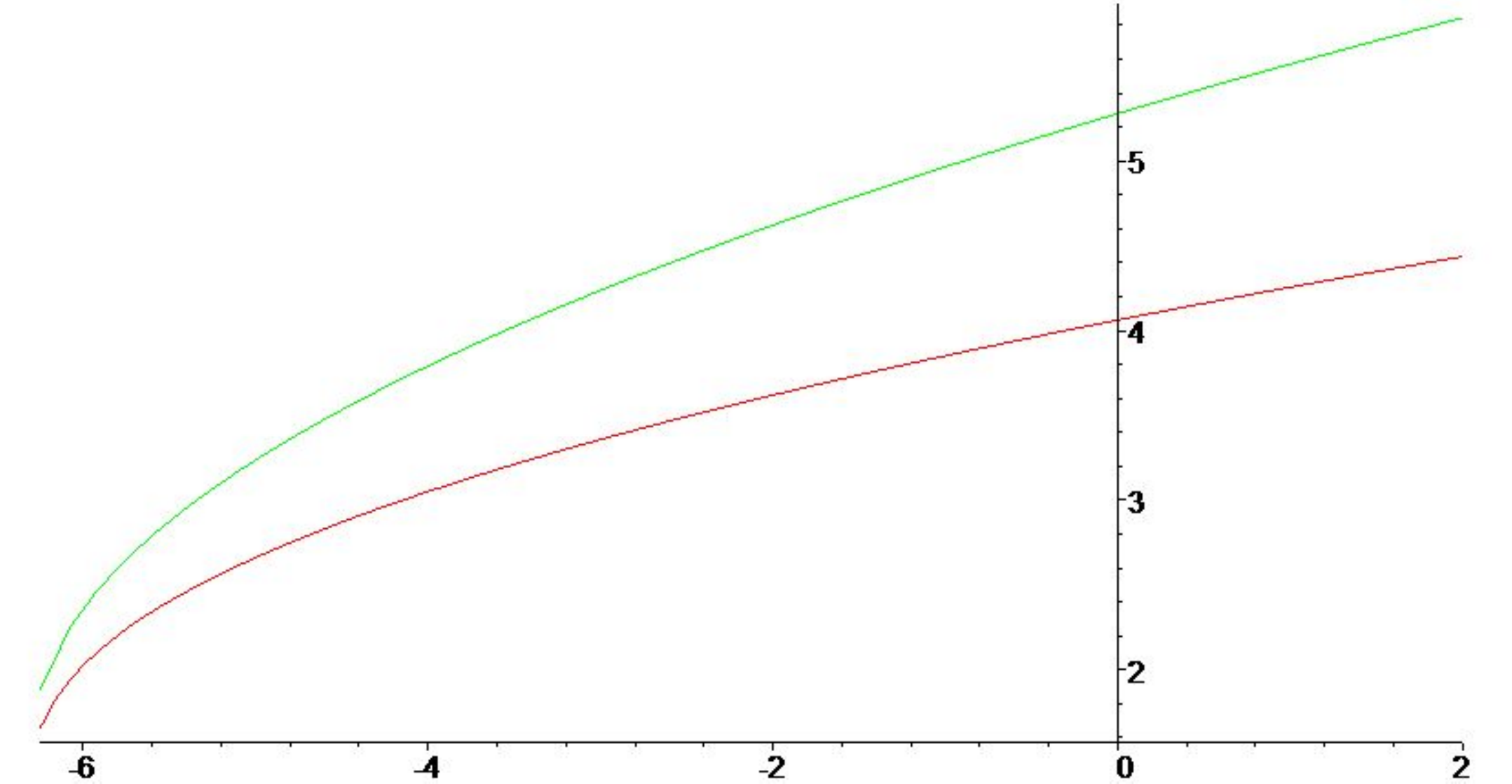}
\end{center}
\caption{\label{mucprobe} Critical values of $\mu q$ vs $m^2$ for scalar condensation in large $q$ limit. The top curve is the critical value for $\mu$ in black hole phase (just above the transition temperature), while the bottom curve is the critical $\mu$ in low temperature phase.}
\end{figure}

As in the low temperature phase, we find a critical value $\mu_c = F(m^2)$  (or, restoring temperature dependence, $\mu_c = {T \over T_c} F(m^2)$) for each choice of $m^2$, above which there is another solution with nonzero $\psi$. This critical $\mu$ may again be determined by a linearized analysis, from which we obtain the equation
\[
\psi'' + \left( {6 r^5 -1  \over r(r^5 - 1)} \right) \psi'  + \left( {\mu^2 (r^3-1)^2 \over  r^4(r^5-1)^2} - { m^2 r^3 \over r^5 -1}\right) \psi =0 \; .
\]
We can set $\psi(1)=1$ without loss of generality, and this requires
\[
\psi'(1) = {m^2 \over 5} \; .
\]
These can be solved numerically to find $F(m^2)$, and our results (with the low temperature results) are plotted in figure \ref{mucprobe}.

A sample phase diagram, for the case $m^2 = -6$ is shown in figure \ref{phaseq}.

\subsection{Order of phase transitions in the probe limit}

To complete this section, we verify analytically that the action for solutions with scalar field in the probe limit is always less than the corresponding unperturbed solution. In this limit we neglect the gravity back reaction of the gauge fields and scalar. The on-shell action in this approximation is given by
\bea
 \frac{S}{T^d} &=&\int d^{d+1}x\ \sqrt{-g}g^{tt}g^{rr}\frac{A_t'^2}{2} \; .
\label{eqn:newform}
\eea
We have used the fact that the scalar action is quadratic and vanishes on-shell once the boundary value of scalar is kept to zero \cite{Arean:2010zw}. Writing the action in this simple form gives us information about the relative free energy of the different phases.

The solution for $A_t$ in the superconducting phase may be written as
\bea
A^{S}_t=A^{0}_t + \delta A_t\,,
\eea
where $\delta A_t \rightarrow 0$ in the IR region of the bulk and near the boundary. $A^{0}_t$ is the value of $A_t$ in the normal phase. Then, from eq. (\ref{eqn:newform}) we get
\bea
\frac{S_{new}}{T^d V} &=& \frac{S_{old}}{T^d V}+ 2 \int dr \sqrt{-g} g^{rr} g^{tt} \partial_r A^{0}_t \partial_r(\delta A_t)+\int \sqrt{-g} g^{rr} g^{tt} \frac{(\delta A_t)'^2}{2} dr \; .
\eea
The cross term between $A^{0}_t$ and $\delta A_t$ vanishes after integrating by parts and then using the eom of $A^{0}_t$. Hence
\be
\delta S= S_{new}-S_{old} = (T^d V) \int \sqrt{-g} g^{rr} g^{tt} \frac{(\delta A_t)'^2}{2} dr < 0  {\rm as } g^{tt}<0.
\label{eqn:at}
\ee
Therefore if a phase with non-trivial scalar condensate exists it will always have a lower free energy than the normal phase and the associated transition will be of second order.

The introduction of gravity may give rise to a positive term in the on-shell action and the nature of phase transition may change.

\section{Critical $\mu$ for solutions with infinitesimal charged scalar}

To find the critical $\mu$ at which solutions with infinitesimal scalar field exist, we find the value of $\mu$ for which the linearized scalar equation about the appropriate background admits a solution with the right boundary conditions at infinity.

At low temperatures, this gives (setting $r_0=1$)
\beas
& \psi'' + \left({g' \over g} + {4 \over r}  \right) \psi'+ {1 \over g} \left({q^2 \phi^2 \over r^2}-m^2\right) \psi \;, \cr
&g(r) = r^2 - {1 \over r^3} \;, \qquad \phi = \mu \; ,
\eeas
while for the RN black hole background (setting $r_+=1$) we have
\beas
&\psi'' + \left({g' \over g} + {4 \over r}  \right) \psi'+ {1 \over g} \left({q^2 \phi^2 \over g}-m^2\right) \psi \;,\cr
&g(r) = r^2 - \left(1 + {3 \mu^2\over 8} \right){1 \over r^3} + {3 \mu^2 \over 8 r^6} \;, \cr &\phi = \mu \left( 1 - {1 \over r^3} \right) \; .
\eeas
More general values of $r_0$ or $r_+$ can be restored by the scaling symmetry.

\begin{figure}[t]
\begin{center}
\includegraphics[width=0.7\textwidth]{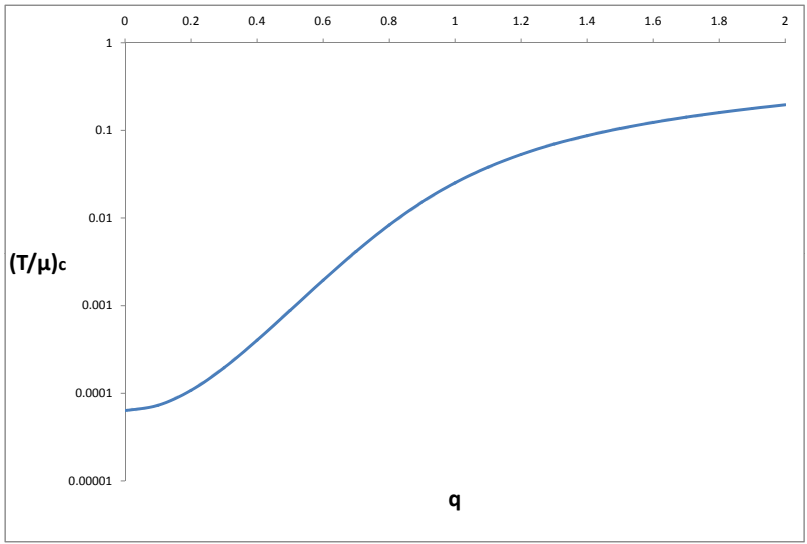}
\end{center}
\caption{\label{Tmuvsq} Critical $T/\mu$ vs charge $q$ for condensation of $m^2 = -6$ scalar field in Reissner-Nordstrom background.}
\end{figure}

For $m^2=-6$, we find a critical value of $\mu$ in the low-temperature case given by $\mu_{low} q = 5.089/(2 \pi R)$. At high temperatures, the critical solutions exist $T/\mu$ when has a critical value as plotted in figure \ref{Tmuvsq}.

\end{document}